# Understanding Mental Health Content on Social Media and It's Effect Towards Suicidal Ideation

Mohaiminul Islam Bhuiyan, Nur Shazwani Kamarudin*, Nur Hafieza Ismail

Faculty of Computing, Universiti Malaysia Pahang Al-Sultan Abdullah, Pekan, Pahang, Malaysia

*Abstract*—The study "Understanding Mental Health Content on Social Media and Its Effect Towards Suicidal Ideation" aims to detail the recognition of suicidal intent through social media, with a focus on the improvement and part of the machine learning (ML), deep learning (DL), and natural language processing (NLP). This review underscores the critical need for effective strategies to identify and support individuals with suicidal ideation, exploiting technological innovations in ML and DL to further suicide prevention efforts. The study details the application of these technologies in analyzing vast amounts of unstructured social media data to detect linguistic patterns, keywords, phrases, tones, and contextual cues associated with suicidal thoughts. It explores various ML and DL models like SVMs, CNNs, LSTM, neural networks, and their effectiveness in interpreting complex data patterns and emotional nuances within text data. The review discusses the potential of these technologies to serve as a life-saving tool by identifying at-risk individuals through their digital traces. Furthermore, it evaluates the real-world effectiveness, limitations, and ethical considerations of employing these technologies for suicide prevention, stressing the importance of responsible development and usage. The study aims to fill critical knowledge gaps by analyzing recent studies, methodologies, tools, and techniques in this field. It highlights the importance of synthesizing current literature to inform practical tools and suicide prevention efforts, guiding innovation in reliable, ethical systems for early intervention. This research synthesis evaluates the intersection of technology and mental health, advocating for the ethical and responsible application of ML, DL, and NLP to offer life-saving potential worldwide while addressing challenges like generalizability, biases, privacy, and the need for further research to ensure these technologies do not exacerbate existing inequities and harms.

## I. INTRODUCTION

As digital technologies permeate and reshape society, social media has emerged as a massive window into the psychological landscape of users. Individual expressions within these platforms can serve as a mirror, presenting critical insights into the mental well-being of the global population. Although revealing various pathologies, a stark reality is reflected that demands urgent attention - suicidal ideation. This affliction continues to take an unconscionable toll, with the World Health Organization (WHO) reporting nearly 800,000 people dying by suicide annually [1]. Shockingly, it is the second leading cause of death for those aged 15-29 years [2]. Many suicides relate to mental health issues like depression, substance abuse, and psychosis. But the causes are complex. Those struggling deserve support [1]. These alarming statistics underscore the human loss and suffering that suicidal behaviors induce worldwide. Clearly, there is a pressing need to earnestly seek more effective strategies for identifying those with suicidal ideation to provide life-saving prediction, prevention and intervention.

Fortunately, advancements in Machine Learning (ML) and Deep Learning (DL) are illuminating a promising path forward, offering technological innovations that could drastically further suicide prevention efforts [3], [4]. The potential of machine learning for suicide prevention is significant, but practical application faces hurdles regarding transparency, ethics, and data quality for which further research is needed [5]. At the core is the ongoing amassment of vast digital traces as we communicate, browse, share, and express ourselves through online conduct. Analyses of resulting "big data" repositories using sophisticated analytical techniques promise to usher impactful progress in decoding and responding to human behaviors and states of mind [6]. Specifically, by applying ML and DL tools to mine social media communications, the recognition of linguistic patterns and semantic complexities associated with suicidal ideation can be realized to an unprecedented degree [7], [8]. Essential human expressions conveyed through unstructured text data can now be computationally elucidated to discern what once remained invisible. This paper aims to comprehensively review the detection of suicidal ideation on social media, focusing on the roles and advancements of machine learning (ML), deep learning (DL), and natural language processing (NLP).

Machine learning, deep learning, and natural language processing are being applied to detect signs of suicidal ideation in social media posts. These technologies can analyze large volumes of unstructured text data to identify linguistic patterns, keywords, phrases, tones, and contextual cues associated with suicidal thoughts. While traditional ML techniques like SVMs and Random Forests can learn predictive rules, they are limited in understanding nuanced semantics and emotions. However, deep learning methods like CNNs and LSTM neural networks, which model complex data patterns, show promise for interpreting broader concepts and context to represent the intricacies of human experience [9], [10], [11]. Though still imperfect, deep learning's ability to comprehend subtext and vulnerability in language offers hope for identifying cries for help and risk factors in a more meaningful way. Advanced NLP and deep learning may enable breakthroughs in computationally decoding the complexities of human expression to uncover suicidal warning signs in text.

Thus, while challenges and ethical quandaries persist, powerful capacities now exist to extract and analyze massive stores of social expressions for life-saving insights. If properly nurtured under responsible development and usage guidelines, AI and learning systems could gain competencies to serve the public good [12]. Though much progress is still sorely needed, thoughtfully designed mechanisms harnessing big data may move within reach to provide rich awareness of mental states, risk and distress... potentially before tragic outcomes occur [5]. With compassion and diligent effort, the possibility of channeling AI's emerging potential shouldn't be readily dismissed. Perhaps computational tools aimed toward understanding hearts and minds could reveal key markers that humans often miss in order to guide care and resources to those facing darkest moments. Lives awaits saving by transforming big data into wisdom and decisive action [13].

This comprehensive review aims to address critical knowledge gaps by thoroughly analyzing existing research on detecting suicidal ideation through social media using machine learning (ML) and deep learning (DL). It will scrutinize the specific methodologies, tools, and techniques utilized in recent studies, assessing their real-world effectiveness, limitations, and ethical considerations. The review traces the evolution of ML and DL in this application, highlighting advancements that show promise while surfacing areas needing additional exploration and development. Questions around generalizability, biases, and privacy represent just some emerging issues that require elucidation if these technologies are to progress responsibly.

Ultimately, the significance of synthesizing current literature lies not merely in its academic contribution, but also in its potential to meaningfully inform practical tools and suicide prevention efforts. By consolidating empirical insights around the capabilities and development needs of ML and DL for suicide risk detection, this review seeks to guide beneficial innovation of reliable, ethical systems. Such systems could enable early intervention, connecting vulnerable individuals with support and resources.

This research synthesis aims to navigate the intersections of technology and mental health by evaluating recent studies, distilling knowledge, and charting an informed path forward. Applied ethically and responsibly, these emerging analytical methods may offer life-saving potential across populations worldwide. But vigilance around limitations and directing further research is required to ensure applications counter, not exacerbate, existing inequities and harms. This review confronts these multifaceted issues to support deliberate, compassionate translation of scientific discoveries into societal solutions. The primary contributions of this paper are:

- Survey key computational techniques for social media-based screening of suicidal ideation.

- Evaluate the effectiveness of combining sentiment analysis with machine learning on benchmark suicide risk assessment tasks.

- Examine strengths and limitations of social media data sources like Twitter, Reddit, Facebook and various mental health forums.

- Discuss ethical implications including privacy, stigma, and duty of care when analysing social media content.

- Synthesize critical directions and opportunities for impactful research at the intersection of NLP, machine learning, mental health, and suicide prevention.

Advanced analytical methods offer capabilities to decode warnings, risks and signs of suicidal ideation expressed through digital traces. This review article will present a comprehensive analysis of the current state of the art and research trajectory in this critical domain. It will scrutinize in depth the methodologies, tools and techniques being applied to social media data. Key dimensions evaluated will encompass demonstrable performance, advantages, limitations and ethical considerations of using machine learning and AI to detect patterns of mental health distress. The overarching aim is to consolidate current knowledge regarding the promises and perils of computationally surveilling the social landscape with techniques that peer deeply into the collective psyche through the lens of big data... potentially illuminating who requires help and enabling intervention with greater acuity. What is at stake warrants continued exploration to thoughtfully harness such technology for the greatest good.

The remainder of this paper is structured as follows: Section II reviews related work on the detection of suicidal ideation from social media. Section III discusses the evolution of methodologies and details current techniques, including data handling and model applications. Finally, Section IV summarizes insights and outlines future research directions.

## II. Previous Works

The detection and analysis of suicidal ideation through social media has gained increasing attention in the field of mental health research. This section provides an overview of the existing literature, outlining the evolution of methodologies and the various approaches used in the detection of suicidal ideation.

In embarking on an analysis of recent studies focused on the detection of suicidal ideation through social media using machine learning (ML), deep learning (DL) and Natural Language Processing (NLP) techniques, we delve into a domain of computational psychiatry that has shown both promising advancements and encountered significant challenges. This analysis will synthesize key findings from various research efforts, evaluating their methodologies, effectiveness, and broader implications within the public health context.

Table I represents a comparison table featuring key details from some of the previous studies on suicidal ideation detection via social media.

The growing prevalence of mental health issues, particularly suicidal ideation, and its manifestation on social media platforms, presents an urgent need for innovative monitoring and intervention strategies. Studies such as those conducted by Samer Muthana et al. [30], Arunima Roy et al. [31], and Swati Jain et al. [32], have explored the utilization of platforms like Twitter, employing techniques ranging from sentiment analysis to complex neural network architectures.

These studies highlight the potential of ML and DL in deciphering the nuanced language of distress and suicidal thoughts expressed in social media posts.

TABLE I.          COMPARISON OF STUDIES ON SUICIDAL IDEATION DETECTION

| Study Reference | Year | Platform | ML/DL Techniques Used | Dataset Size | Key Features Analyzed | Results (e.g., Accuracy, Precision, Recall) | Limitations |
|---|---|---|---|---|---|---|---|
| Jorge Parraga-Alva et al. [14] | 2019 | Various | Unsupervised learning | 102 texts | Suicide message categorization | Avg rate: 79% - 87% | -small corpus -lack of multiple categories |
| Pratyaksh Jain et al. [15] | 2022 | Reddit | Naïve Bayes, SVM, Logistic Regression | 60,000 data points | Depression, Suicidal behavior | Acc: 74.35% - 77.29%, F1: ~0.77 | -still room for improve the predictive model |
| Syed Tanzeel Rabani et al. [16] | 2020 | Twitter | Random Forest | 4,266 tweets | Suicidal ideation | Acc: 98.5%, Precision: 98.7%, Rec: 98.2% | -no direct communication or intervention |
| Seunghyong Ryu et al. [17] | 2019 | KNHANES | Random forest | 5,773 subjects | Suicide ideation | AUC: 0.947, Acc: 88.9% | -One ML algorithm -SMOTE used to overcome class imbalance problem |
| Ning Wang et al. [18] | 2021 | Social media | KNN, SVM, C-Attention network | CLPsych 2021 dataset | Suicide attempts prediction | Best F1 score: 0.737 (6 months prior) | - Traditional ML methods better in some metrics |
| Samh J. Fodeh et al. [19] | 2019 | Twitter | LSA, LDA, NMF and Decision Tree and K-means Clustering | 12,066 Tweets | Suicide risk factors | Precision: 0.844, Sensitivity: 0.912 | -Biased Data Selection -Missing Twitter Metadata -Ignored Ground Truth |
| Shi Ru Lim et al. [20] | 2023 | Twitter | SVM, Decision Tree, Naïve bayes | 16,158 Tweets | Mental health disorders prediction | Acc: SVM: 99.80% (training), 98.43% (testing) | -Larger dataset needed |
| Tianlin Zhag et al. [21] | 2021 | Online | Transformer RNN | 659 samples | Suicide notes identification | 94.9% recall, 94.9% F1 score, 95% precision | -Explore linguistic -psychological features |
| Ayaan Haque et al. [22] | 2021 | Reddit, IMDB large movie dataset | SDCNL, BERT, Sentence-BERT, GUSE | 1,895 posts, 50,000 reviews | Depression vs suicidal ideation classification | Strong performance in classification | -Evaluate on other datasets |
| Yaakov Ophir et al. [23] | 2020 | Facebook | ANN models (STM, MTM) | 83,292 posts of 1002 users | Detect suicide risk | AUC: .697 to .746 (MTM) | -practical detection tools for suicide risk |
| Tadesse et al. [7] | 2019 | Reddit | LSTM-CNN, word embedding | Posts from Suicide Watch forum | Suicidal ideation detection | Acc: 93.8% (LSTM-CNN) | -data deficiency -annotation bias |
| Prasadith Buddhitha et al. [24] | 2019 | WASSA-2017, CLPsych-2015 | Deep Neural Network | - | Depression, PTSD detection | AUC: >90% (multi-channel CNN) | -insufficient unstructured data -lack of intervention |
| Diniz et al. [25] | 2022 | Twitter | ML, DL, BerTimbau Large | 3788 instances | Suicidal ideation from text | Acc: 0.955, F1: 0.954 | -inability to detect typos |
| Aldhyani et al. [26] | 2022 | Reddit | CNN-BiLSTM, XGBoost | 232074 samples | Textual and LIWC based features from post | Acc: 95% (Textual) Acc: 84.5% (LIWC) | -variations in performance |
| Renjith et al. [27] | 2022 | Reddit | LSTM-Attention-CNN | 55,680 posts | Suicidal ideation in text | Acc: 90.3%, F1 Score: 92.6% | - |
| Kholifah et al. [28] | 2020 | Twitter | LSTM Deep Learning model | - | Level of depression | Acc: 70.89%, Precision: 50.24%, Recall: 70.89% | -accuracy of the classification results can be improved |
| Cao et al. [29] | 2022 | Microblog, Reddit | Deep neural networks integrated with knowledge graph | 7,329 subjects | Personal factors related to suicidal ideation | Acc: >93% | -Data lacking, noisy abundance |

However, this promising avenue is not without its complexities. The critical evaluation of these studies reveals a common challenge – the balance between the technical efficacy of detection algorithms and the ethical implications of privacy and data usage. For instance, while studies like those by Pratyaksh Jain et al. [15] on Reddit and Syed Tanzeel Rabani et al. [16] on Twitter demonstrate high accuracies in detecting suicidal ideation, they also raise questions about the representativeness of their datasets and the ethical management of sensitive personal information.

Moreover, the generalizability of these models across different demographics and social media platforms remains a significant concern. Studies often use datasets that may not adequately represent the broader population, leading to potential biases in prediction models. This limitation is evident in the work of researchers like Jorge Parraga-Alva et al. [14] and Hannah Yao et al. [33], who have tried to categorize and understand the complexity of suicide-related messages across various platforms. Seunghyong Ryu et al. [17] and Tatiana Falcone et al.'s [34] studies offer unique insights into suicide ideation detection in specific contexts. Ryu et al. [17] used data from the Korea National Health & Nutrition Examination Survey to develop predictive models for identifying individuals at risk of suicide, demonstrating the use of machine learning in a clinical or community setting. Falcone et al.'s [34] research on open-source digital conversations among people with epilepsy provides an interesting perspective on how machine learning can uncover specific concerns and struggles related to suicidality in distinct patient groups.

Jianhong Luo et al. [35] and Ning Wang et al. [18] both utilized Twitter data for their research but with different focuses. Luo et al. [35] examined temporal patterns of potential suicidal ideations, using topic modeling to identify latent suicide topics, while Wang et al. [18] developed models for predicting suicide attempts based on social media posts. These studies reveal the complexity and multifaceted nature of suicidal behavior as expressed on social media, and the potential of machine learning in capturing these nuances. John Torous et al. [5] and Gen-Min Lin et al. [36] addressed the integration of technology in suicide prevention and prediction. Torous et al. discussed the potential of tools like text messaging, smartphone apps, and machine learning algorithms in suicide prevention, while Lin et al. focused on predicting suicide ideation in military personnel using various machine learning techniques. These studies illustrate the diverse applications of machine learning and technology in different contexts and populations for suicide prediction and prevention.

Xingyun Liu et al. [37], ML Tlachac et al. [38], and E. Rajesh Kumar et al. [39] explored novel approaches in suicide ideation detection. Liu et al. assessed the feasibility of a proactive suicide prevention approach on social media, while Tlachac et al. investigated the use of smartphone-based communication for passive screening of suicidal ideation. Kumar et al. focused on Twitter data to establish an early warning system for suicidal thoughts detection. These studies highlight innovative applications of machine learning in real-time and proactive detection of suicide risk.

Atika Mbarek et al. [40] and Samh J. Fodeh et al. [19] emphasized the detection of suicidal profiles and factors contributing to suicide risk on Twitter. Mbarek et al. developed a machine learning-based approach for detecting suicidal profiles, while Fodeh et al. used machine learning algorithms to identify factors related to suicide risk. These studies underscore the potential of machine learning in profiling and understanding the underlying factors of suicide risk on social media platforms. Shi Ru Lim et al. [20] and Ibrahim et al.'s [41] studies focused on the prediction of mental health disorders using machine learning techniques. Lim et al. applied various algorithms like SVM and Naive Bayes to predict mental health issues based on Twitter data, while Ibrahim et al. detected PTSD symptoms through Twitter postings. Both studies demonstrate the broader application of machine learning in mental health beyond suicide prevention.

Kamarudin et al.'s [42], [43] two studies expanded the scope of machine learning in understanding mental health challenges through social media data. Their first study employed machine learning and natural language processing to dissect data from virtual communities, while their second study focused on linguistic analysis of online mental health communities. These studies highlight the rich potential of social media data in increasing mental health awareness and informing policy decisions. Kamarudin et al. [44] conducted another study exploring the role of Reddit in providing support to individuals dealing with rape and sexual abuse. Using natural language processing, they analyzed the diversity of topics in responses, including emotional support, relationships, therapy, and legal advice. The findings highlighted the comprehensive nature of support on Reddit, differing from traditional physical-world responses. The study suggests future research involving detailed surveys to understand the perspectives of both support seekers and providers.

Farsheed Haque et al. [45] and Shini Renjith et al. [27] both employed advanced machine learning models for detecting suicidal ideation in social media posts. Haque et al. used Transformer models for text classification, while Renjith et al. developed an ensemble deep learning technique for fast and accurate detection of suicidal ideation. These studies showcase the advancements in machine learning algorithms for suicide ideation detection.

Tianlin Zhang et al. [21], Ayaan Haque et al. [22], and Yaakov Ophir et al. [23] further contributed to the field by developing specialized models and algorithms for suicide and depression classification. Zhang et al. developed a transformer-based deep learning model for identifying suicide notes online, Haque et al. proposed an unsupervised label correction method for classifying depression and suicidal ideation, and Ophir et al. also focused on distinguishing between depression and suicidal ideation using advanced word embedding models. These studies represent the cutting-edge of machine learning applications in mental health.

Michael Mesfin Tadesse et al. [7] and Ramit Sawhney et al. [8] utilized deep learning techniques for detecting suicide ideation in social media. Tadesse et al. developed an LSTM-CNN model for early detection of suicide ideation, while Sawhney et al. compared various deep learning architectures for suicidal ideation detection. These studies reflect the ongoing evolution and refinement of machine learning techniques in the context of suicide prevention. Prasadith Buddhitha et al. [24] proposed a unique multi-task learning approach with a multi-channel CNN for detecting depression and PTSD. Their approach of incorporating emotional patterns identified by clinical practitioners demonstrates an innovative integration of clinical insights with machine learning.

These studies collectively represent a significant advancement in the use of machine learning and natural language processing techniques to detect, predict, and understand suicide ideation and mental health issues through social media platforms. The diversity in methods, from supervised and unsupervised learning to deep learning and ensemble techniques, highlights the versatility and potential of these technologies in addressing the critical issue of mental health. Moreover, the focus on different populations, languages, and social media platforms underscores the need for tailored approaches that consider the unique aspects of each context and population group. As machine learning continues to evolve, it holds immense promise for enhancing our understanding and prevention of mental health issues, particularly in the realm of suicide prevention. The analysis of these studies also underscores the importance of interdisciplinary collaboration in this field. The integration of insights from psychology, computer science, and ethical considerations is crucial for the development of reliable, effective, and ethically sound detection systems.

This critical analysis aims to provide a comprehensive overview of the current state of research in detecting suicidal ideation through social media. It will highlight the innovations and strengths of recent studies while also addressing the significant challenges and limitations that need to be overcome to harness the full potential of ML and DL in this critical area of public health.

### III. EARLY STYLES AND EVOLUTION

Initial research in the field of suicidal ideation detection was primarily focused on traditional psychological assessments and face-to-face interactions. However, the emergence of social media as a pervasive communication medium opened new avenue for mental health research. Early studies in this domain utilized basic text analysis techniques, relying on keyword searches and simple statistical methods to identify posts indicative of suicidal thoughts. These methods, though pioneering, were limited in their ability to understand the complexities of human language and context.

#### A. Advancement in Machine Learning (ML)

The introduction of machine learning algorithms marked a significant advancement in this field. Machine learning, particularly supervised learning techniques, began to be used to identify patterns and features in social media posts that were indicative of suicidal ideation. These approaches involved training models on annotated datasets where posts were labeled as showing signs of suicidal ideation or not. Commonly used machine learning algorithms in this context included Support Vector Machines (SVM), Decision Trees, and Naive Bayes classifiers.

One of the key challenges in applying machine learning was the need for large, annotated datasets. The creation of such datasets involved manually labeling social media posts, a process that was both time-consuming and subject to human error and bias. Despite these challenges, machine learning algorithms proved to be significantly more effective than earlier methods, as they could capture a wider range of linguistic and semantic features.

#### B. The Role of Natural Language Processing (NLP)

Natural language processing (NLP) emerged as a critical tool in enhancing the effectiveness of machine learning models. NLP techniques allowed for more sophisticated analysis of text data, enabling the extraction of features such as sentiment, thematic elements, and linguistic structures. The use of NLP in conjunction with machine learning algorithms led to more nuanced and accurate detection of suicidal ideation. Techniques such as tokenization, stemming, and lemmatization were employed to preprocess the data, making it more amenable for machine learning models. Sentiment analysis, a subset of NLP, was particularly useful in identifying the emotional tone of posts, which is a crucial aspect in detecting suicidal ideation.

#### C. The New Frontier of Deep Learning (DL)

The advent of deep learning brought about a paradigm shift in the detection of suicidal ideation on social media. Deep learning models, particularly neural networks, were capable of modeling complex patterns in large datasets. Convolutional Neural Networks (CNNs) and Recurrent Neural Networks (RNNs), including their variants like Long Short-Term Memory (LSTM) networks, started being used extensively. These models excelled at understanding the contextual nuances and varied expressions of language, making them particularly effective for this task.

Deep learning models were trained on vast amounts of unstructured social media data, learning to identify subtle patterns and linguistic cues associated with suicidal ideation. These models outperformed traditional machine learning algorithms in many aspects, particularly in their ability to understand the context and semantic meaning of text. The application of deep learning, however, required substantial computational resources and large, well-annotated datasets.

#### D. Integration of ML, DL and NLP

The integration of machine learning, deep learning, and NLP represents the current state-of-the-art in suicidal ideation detection on social media. This integrated approach leverages the strengths of each of these technologies, offering a more robust and accurate detection mechanism. Machine learning algorithms provide the basis for pattern recognition, deep learning models add an understanding of contextual and semantic nuances, and NLP techniques enable the effective processing and analysis of textual data.

A comprehensive comparison table focusing on suicidal ideation using NLP, Machine Learning, and Deep Learning including various models, their features, performance metrics, and other relevant details is given in Table II:

TABLE II. Comprehensive Comparison Table of NLP, ML, DL Models for Suicidal Ideation Detection

| Feature/Model | Traditional ML Models | Deep Learning Models | NLP Techniques |
|---|---|---|---|
| Model Examples | - SVM (Support Vector Machine)<br>- Decision Trees<br>- Random Forest | - CNN (Convolutional Neural Network)<br>- RNN (Recurrent Neural Network)<br>- LSTM (Long Short-Term Memory) | - Sentiment Analysis<br>- Topic Modeling<br>- Word Embeddings |
| Data Requirements | - Structured data<br>- Limited data size | - Large datasets<br>- Unstructured data | - Large text corpora<br>- Contextual data |
| Processing Time | - Generally faster<br>- Less computational resources | - Longer due to complexity<br>- High computational resources | - Varies based on technique and data size |
| Accuracy | - Moderate (depending on the dataset) | - High (especially with large and complex datasets) | - Depends on the complexity of the language and context |
| Interpretability | - High (easier to understand model decisions) | - Low (often referred to as "black box" models) | - Moderate (varies with techniques) |
| Suitability for Short Text (e.g., Tweets) | - Less effective due to limited context | - More effective with context preservation techniques | - Effective with appropriate preprocessing |
| Common Challenges | - Overfitting on small datasets<br>- Limited in capturing complex patterns | - Requires large training datasets<br>- Overfitting risks | - Handling of sarcasm, idioms, and contextual meanings |
| Preprocessing Needs | - Feature selection<br>- Data cleaning | - Data normalization<br>- Sequence padding (for RNN, LSTM) | - Tokenization<br>- Stop word removal<br>- Lemmatization/Stemming |
| Application in Suicidal Ideation Detection | - Effective in structured, labeled data<br>- Quick initial assessments | - High accuracy in pattern recognition<br>- Effective in large-scale social media data analysis | - Essential for understanding linguistic nuances<br>- Contextual sentiment analysis |

*E. Challenges and Ethical Considerations*

Despite the advancements in technology, there are significant challenges in this field. The ethical considerations of privacy and consent are paramount. The use of social media data for mental health research raises questions about the privacy of individuals and the potential for misuse of sensitive information. Ensuring that research in this area is conducted with strict ethical guidelines is crucial.

Another challenge is the accuracy and generalizability of these models. Suicidal ideation is a complex and deeply personal phenomenon, and its expression can vary greatly among individuals. Models trained on specific datasets may not generalize well to broader populations. Additionally, the risk of false positives and false negatives in detection is a concern, as it can have serious implications for the individuals involved.

The literature in the field of suicidal ideation detection on social media using machine learning, deep learning, and NLP indicates a rapidly evolving landscape. From basic text analysis to sophisticated deep learning models, the methodologies have advanced significantly. The integration of these technologies offers promising avenues for early detection and intervention in mental health crises. However, the challenges of data privacy, ethical considerations, and the need for accurate, generalizable models remain areas that require ongoing research and attention.

*F. Commonly Followed Methods*

This section outlines the methodologies and techniques employed in the detection of suicidal ideation on social media, focusing on dataset collection, preprocessing, feature extraction, and the application of machine learning, deep learning, and NLP models.

Following taxonomy represents various approaches and tools used in the studies. Each method or technique is nested within its broader category, illustrating the hierarchical relationship between them:

*1) Dataset collection:* One of the initial steps in studying suicidal ideation using social media data is the collection of relevant datasets. This process involves selecting social media platforms (such as Twitter, Facebook, Reddit, etc.) and extracting posts that potentially indicate suicidal thoughts or behaviors. The selection criteria for these posts often involve keyword searches, using terms commonly associated with suicidal ideation.

The complexity of dataset collection is amplified by ethical considerations, such as the privacy and consent of social media users. Researchers must navigate these challenges while ensuring that the data is representative and unbiased. Additionally, the dynamic nature of social media, where content is continually created and modified, poses a challenge in creating stable and reliable datasets.

*2) Preprocessing:* Once the datasets are collected, the next crucial step is preprocessing. This stage involves cleaning and preparing the data for analysis, which is pivotal in NLP and machine learning applications.

*3)* Some Common Techniques for Suicidal Ideation Detection

*a) Lexicon based emotion analysis*

*i) NRC Affect Intensity Lexicon:* The NRC Affect Intensity Lexicon is a lexicon that provides real-valued scores of affect intensity for a large set of English words. It's an extension of the NRC Emotion Lexicon and was developed by Saif M. Mohammad, a senior research officer at the National Research Council Canada [46]. The lexicon aims to quantify the intensity of the affect (emotion) evoked by a word. Each word in the lexicon is associated with an affect intensity score for each of eight basic emotions: anger, fear, anticipation, trust, surprise, sadness, joy, and disgust. The score ranges from 0 (no association with the emotion) to 1 (strong association).

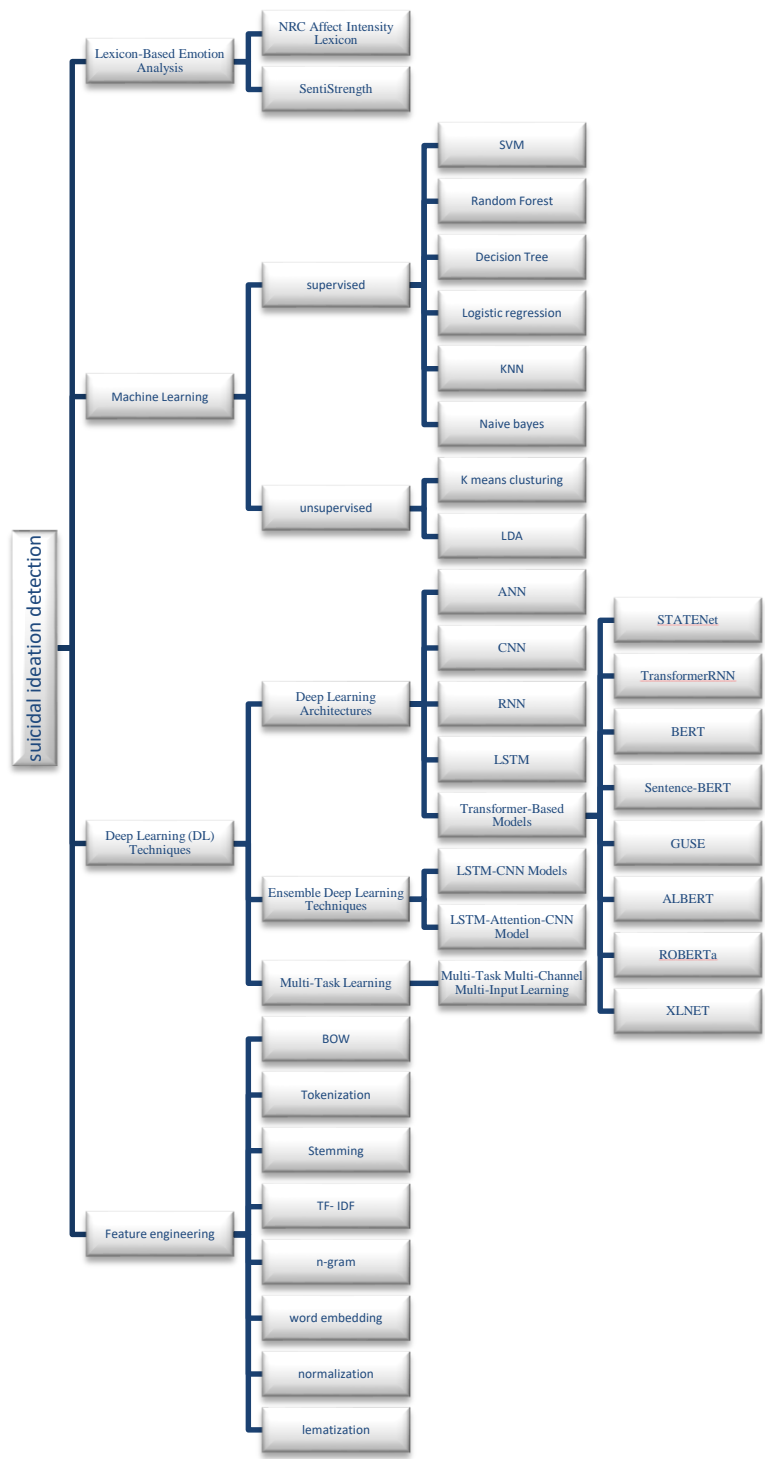

Fig. 1. Hierarchical taxonomy of methodologies for detecting suicidal ideation on social media.

It's widely used in sentiment analysis, social media monitoring, customer feedback analysis, and other areas of natural language processing (NLP) where understanding emotional content is important [47].

*ii) SentiStrength:* SentiStrength is a sentiment analysis tool designed to decipher the sentiment strength of texts, particularly short informal text found on social media platforms. It is based on a lexicon approach, which means it relies on a dictionary of sentiment-related words, each tagged with sentiment strength scores [48]. SentiStrength can be fine-tuned for specific domains (e.g., software engineering) [49] by adjusting its dictionary to include jargon and terminologies unique to those fields. It can also be adapted to different contexts and languages, recognizing that sentiment expression varies greatly across cultural and linguistic landscapes. It is used to gauge public sentiment on various topics by analysing social media posts on platforms like Twitter and YouTube. Governments and organizations use it to understand public opinion on policies, products, or events. SentiStrength struggles with posts that contain images [50], as its lexical approach does not account for visual sentiment indicators. Like many sentiment analysis tools, SentiStrength may not accurately interpret sarcasm and irony, which are prevalent in social media text. The rapid evolution of online language can outpace the tool's dictionary updates, potentially leading to inaccurate sentiment scores.

SentiStrength stands out for its ability to decipher sentiment in informal, web-based text, making it a valuable tool for researchers and organizations looking to tap into public sentiment.

The heart of suicidal ideation detection lies in the application of machine learning and deep learning models. These models are trained on preprocessed and feature-extracted data to classify social media posts as indicative of suicidal ideation or not.

*b) Supervised machine learning models*

*i) Support Vector Machines (SVM):* Support Vector Machine (SVM) is a robust and versatile supervised machine learning algorithm widely used for classification and regression tasks [51], but primarily known for its applications in classification. The core idea behind SVM is to find the optimal hyperplane that maximizes the margin between different classes in the feature space. For a linear SVM (the simplest form) [52], the decision boundary is a hyperplane, and the goal is to find the hyperplane that leaves the maximum margin from the nearest points of all classes, which are known as support vectors. Mathematically, this can be represented by the equation of the hyperplane:

$$w.x - b = 0 \qquad (1)$$

where $w$ is the weight vector perpendicular to the hyperplane, $x$ represents the input features, and $b$ is the bias term. In more complex scenarios, where classes are not linearly separable, SVM uses kernel functions to transform the input space into a higher-dimensional space [53] where a hyperplane can effectively segregate the classes. This is known as the Kernel trick, allowing SVM to perform non-linear classification.

SVM's strength lies in its versatility and efficiency, particularly in high-dimensional spaces. It's relatively memory efficient as it uses a subset of training points in the decision function (the support vectors), making it both powerful and efficient. The optimization of SVM involves finding the right balance between increasing the margin size and minimizing classification error, controlled by a regularization parameter. Support Vector Machine (SVM) is adept at detecting suicidal ideation from social media data, largely due to its proficiency in managing high-dimensional spaces [53]. By employing techniques like TF-IDF for feature extraction, SVM transforms complex and nuanced social media text into a structured format. It then identifies an optimal hyperplane to differentiate between posts that indicate suicidal ideation and those that don't, maximizing the margin between these categories using key data points, known as support vectors. This approach is particularly effective due to the subtle linguistic cues and contextual nuances present in social media content, though its success is contingent on precise data preprocessing and feature selection to capture the complexities of emotional expressions online.

*ii) Decision tree:* A Decision Tree is a widely used non-parametric supervised learning algorithm used for classification and regression tasks [54]. It models decisions and their possible consequences as a tree-like structure, where each internal node represents a "test" on an attribute, each branch represents the outcome of the test, and each leaf node represents a class label (decision taken after computing all attributes). The paths from root to leaf represent classification rules. In building the decision tree, algorithms like ID3, C4.5, or CART [54] are used to determine how to split the data and construct the tree. These algorithms typically employ measures such as Gini impurity or entropy to choose which feature to split at each step. The Gini impurity for a set is given by:

$$Gini\ Impurity = 1 - \sum_{i=1}^{n} p_i^2 \qquad (2)$$

Where $p_i$ is the probability of an object being classified to a particular class. Alternatively, entropy, a measure of disorder or impurity, is given by:

$$Entropy = -\sum_{i=1}^{n} p_i \log_2(p_i) \qquad (3)$$

In the context of detecting suicidal ideation from social media text data, a Decision Tree algorithm works by learning from the features extracted from the text (such as specific keywords, phrases, sentiment scores, or linguistic patterns) and creating a model that can classify new texts based on these learned patterns [38]. Each node in the tree represents a feature, and the branches represent the decision rules leading to the final classification in the leaf nodes. This approach is particularly suitable for such tasks due to its interpretability; the tree structure allows for easy understanding and tracing of the decision path and reasoning [19]. This transparency is crucial in sensitive applications like mental health monitoring, where understanding the rationale behind a classification (such as why a particular post is flagged as indicative of suicidal ideation) is as important as the classification itself.

*iii) Random forest:* Random Forest is a machine learning technique for classification, regression, and other tasks that operates by constructing a multitude of decision trees at training time and outputting the mode of the classes or mean prediction of the individual trees [55]. It builds upon the concept of bagging, an approach that improves the stability and accuracy of machine learning algorithms by combining multiple models to 'vote' on the final output [56]. Each tree in a Random Forest makes a class prediction, and the class with the most votes becomes the model's overall prediction. The algorithm injects randomness into the model building process, which not only helps in reducing the variance of the model but also prevents overfitting. This is done by randomly selecting subsets of the training data set (with replacement) and subsets of the features used for splitting nodes. The decision tree equation typically involves measures like Gini impurity or entropy [57] to find the best split at each node:

$$Gini\ Impurity = 1 - \sum(p_i)^2 \qquad (4)$$

$$Entropy = -\sum p_i \log_2(p_i) \qquad (5)$$

Where $p_i$ is the proportion of samples that belong to a certain class in a given node.

In the context of suicidal ideation detection from social media text data, Random Forest can be highly effective [7]. It starts with preprocessing and transforming textual data into a numerical format, typically using vectorization methods such as TF-IDF or word embeddings. Each decision tree in the Random Forest then learns from a random subset of these features, creating a diverse set of perspectives for interpreting the data. When new data is inputted, each tree makes a prediction based on its learned patterns, and the final output is decided by a majority vote across all trees. This methodology is particularly suitable for analyzing social media content, as it accommodates the high variability and noise often found in this type of data. The ensemble nature of Random Forest also helps in capturing a wide array of linguistic and contextual indicators associated with suicidal ideation, making it a robust tool for mental health monitoring in digital platforms [16], [17].

*iv) Logistic regression:* Logistic Regression is a statistical method used for binary classification problems, which models the probability of a binary response based on one or more predictor variables [58]. It operates under the principle that the log odds of the probability of an event occurring versus it not occurring is linearly related to the independent variables [59]. This relationship is expressed through the logistic function:

$$\log\left(\frac{p}{1-p}\right) = \beta_0 + \beta_1 X_1 + \beta_2 X_2 + \cdots + \beta_n X_n \qquad (6)$$

Here, $p$ is the probability of the dependent event occurring, $\beta_0$ is the intercept, $\beta_1$, $\beta_2$, ..., $\beta_n$ ) are the regression coefficients, and $X_1, X_2, ..., X_n$ are the independent variables. The output is then transformed using the logistic function (or sigmoid function), ensuring that the output always lies between 0 and 1, making it interpretable as a probability:

$$p = \frac{1}{1 + e^{-(\beta_0 + \beta_1 X_1 + \beta_2 X_2 + \cdots + \beta_n X_n)}} \qquad (7)$$

In utilizing Logistic Regression for identifying suicidal ideation within social media text, a nuanced approach is adopted to process and analyze the content [32]. Initially, the textual data undergoes preprocessing to extract meaningful attributes, which might include the frequency of emotive words, the structure of sentences, or the usage of certain language patterns associated with mental health indicators. Logistic Regression then applies its algorithm to these extracted features, determining the likelihood of a post reflecting suicidal thoughts. The output, given as a probability, offers a graded scale of risk assessment rather than a binary classification, providing a more refined analysis. This method stands out for its straightforward interpretability, essential for mental health monitoring, where understanding the reasoning behind a prediction is crucial. However, the model's effectiveness hinges on the relevance and quality of the features selected, necessitating careful curation to avoid biases or misinterpretations [15], [32]. Such an approach allows Logistic Regression to be a potent tool in the sensitive area of mental health surveillance on digital platforms.

*v) KNN:* K-Nearest Neighbors (KNN) is a simple, yet versatile supervised learning algorithm used for both classification and regression tasks, but it's most utilized for classification [60]. KNN operates on the principle that similar things exist in proximity; in other words, it assumes that similar data points are near to each other. The algorithm doesn't explicitly learn a model; instead, it classifies a new data point based on the majority vote of its 'K' nearest neighbors. The number of neighbors, K, is a critical user-defined parameter, and the distance between data points is calculated using metrics like Euclidean distance, Manhattan distance, or Hamming distance [61]. The Euclidean distance between two points in a plane is given by [62]:

$$d(p,q) = \sqrt{(q_1 - p_1)^2 + (q_2 - p_2)^2 + \cdots + (q_n - p_n)^2} \quad (8)$$

In suicidal ideation detection from social media text using K-Nearest Neighbors (KNN), the algorithm classifies new posts based on linguistic and emotional similarities with existing posts [18]. After transforming text data into numerical features, KNN identifies the 'K' closest posts in this feature space, determining the new post's category based on the prevalence of categories (indicative of suicidal ideation or not) among these nearest neighbors. This method effectively captures subtle patterns in personal expression [41], with the choice of 'K' and the nature of features used being crucial for accuracy. The model's reliance on direct comparisons with known cases makes it uniquely suited for discerning nuanced expressions of mental states in social media, provided the training data is diverse and representative.

*vi) Naïve bayes classifiers:* Naïve Bayes is a probabilistic machine learning model based on applying Bayes' Theorem with the assumption of independence among predictors [63]. It's particularly popular for text classification tasks due to its simplicity and effectiveness, even with high-dimensional data. The fundamental equation is Bayes' Theorem [64]:

$$P(A/B) = \frac{P(B/A)P(A)}{P(B)} \qquad (9)$$

In this context, $A$ and $B$ represent different events, where $P(A/B)$ is the probability of $A$ given $B$, $P(B/A)$ is the probability of $B$ given $A$, $P(A)$ and $P(B)$ are the probabilities of observing $A$ and $B$ independently of each other.

For suicidal ideation detection in social media texts, Naive Bayes works by first transforming text data into a feature vector (often using techniques like bag-of-words or TF-IDF). Each word or phrase in the text contributes independently to the probability that the message reflects suicidal ideation. The model then uses these probabilities, derived from the training data, to predict whether new texts suggest suicidal thoughts [15], [20]. This method's efficiency in handling large volumes of text and its ability to quickly classify new data make it suitable for real-time monitoring of social media platforms. However, the assumption of independence among words can sometimes oversimplify the linguistic complexities of human communication, particularly in the nuanced context of mental health discussions. Despite this, Naive Bayes remains a popular choice due to its ease of implementation and its proficiency in processing and classifying large datasets efficiently.

*c) Unsupervised machine learning models*

*i) K-Means Clustering:* K-Means Clustering is an unsupervised learning algorithm that partitions a dataset into k clusters by minimizing the intra-cluster variance [65]. It does so by iteratively assigning data points to the nearest cluster centroid and updating the centroid to the mean of the points in the cluster. The algorithm's objective is to minimize the sum of squared distances between points and their respective cluster centroids, expressed as:

$$J = \sum_{j=1}^{k} \sum_{i=1}^{n} \left\| x_i^{(j)} - c_j \right\|^2 \qquad (10)$$

Here, $\left\| x_i^{(j)} - c_j \right\|^2$ is the Euclidean distance between a data point $x_i^{(j)}$ and the cluster centroid $c_j$, $n$ is the number of data points in cluster $j$, and $k$ is the number of clusters.

For suicidal ideation detection in social media text, K-Means can group posts into clusters based on textual similarities, using features like TF-IDF vectors. These clusters can then be analyzed to identify common themes or expressions indicative of suicidal thoughts [19]. This clustering approach helps in organizing large, unlabeled datasets into meaningful categories, facilitating the identification of mental health concerns. However, the effectiveness of K-Means in this context heavily relies on the choice of k and the initial centroid selection, alongside robust preprocessing of the text data.

*ii) Latent Dirichlet Allocation (LDA):* Latent Dirichlet Allocation (LDA) is a popular unsupervised learning algorithm in natural language processing, primarily used for topic modeling [66]. It assumes that documents are mixtures of topics and that topics are mixtures of words. This generative model allows it to discover hidden thematic structures in large collections of text documents. LDA represents each document as a distribution over topics and each topic as a distribution over words.

For detecting suicidal ideation in social media texts, LDA helps identify topics that may signify mental distress. It processes text data to extract latent topics, potentially flagging those aligned with suicidal thoughts. This method is valuable for exploratory analysis, identifying linguistic patterns related to mental health [19], [43]. However, the effectiveness of LDA depends on accurate text preprocessing and parameter tuning, requiring careful interpretation in the sensitive context of mental health monitoring.

*d) Deep learning models*

*i) Artificial Neural Networks (ANN):* Artificial Neural Networks (ANNs) are computational models inspired by the human brain's structure and function, particularly effective in pattern recognition and predictive modeling [67]. An ANN consists of layers of interconnected nodes or neurons, each node in one layer connected to several others in the next layer. The basic operation in a neuron involves weighted input summation and a non-linear activation function [68]. Mathematically, the output $y$ of a neuron can be described by the equation,

$$y = f(\sum_{i=1}^{n} w_i x_i + b) \qquad (11)$$

where $x_i$ are the inputs, $w_i$ are the weights, $b$ is the bias, and $f$ is the activation function, like a sigmoid or ReLU function. The network learns to model complex relationships by adjusting these weights and biases based on the input data.

For suicidal ideation detection from social media text, ANNs can be employed to analyze and classify text data [23]. The process involves preprocessing the text (like tokenization, stemming, and vectorization) to transform it into a format suitable for the ANN. The network then processes this data through its layers, learning to identify patterns and linguistic cues that are indicative of suicidal ideation. This might include specific keywords, phrases, or the overall sentiment of the text. The final layer of the network, typically a SoftMax layer, classifies the text based on the learned patterns, indicating whether it likely contains suicidal ideation. The performance of ANNs in this application largely depends on the network's depth and complexity, the quality of the training data, and the model's capacity to capture subtle nuances in the language that may suggest suicidal thoughts or tendencies.

*ii) Convolutional Neural Networks (CNN):* CNN, or Convolutional Neural Networks, are a class of deep neural networks commonly used in image processing [69] but also effective in text classification [70], particularly for feature extraction [71]. CNNs are particularly effective due to their ability to learn spatial hierarchies of features through the use of convolutional layers. A basic CNN structure includes an input layer, convolutional layers, pooling layers, fully connected layers, and an output layer [72]. The convolutional layers apply a convolution operation to the input, passing the result to the next layer. This can be represented by the equation:

$$Output = Activation(Weights.Input + Bias) \qquad (12)$$

where Activation is a non-linear function like ReLU. Pooling layers reduce the spatial size of the convolved features to decrease the computational power required. Finally, fully connected layers combine these features to classify the input into various categories.

For suicidal ideation detection from social media text, CNNs can be utilized to analyze and classify textual data [27]. The process begins with the collection and preprocessing of social media texts, which involves cleaning, normalization, and tokenization to convert text into a form that can be fed into a CNN. CNN then learns to identify patterns and features in the text indicative of suicidal ideation, such as specific keywords, phrases, or linguistic patterns. This involves multiple layers of convolution and pooling to extract and condense the relevant features from the text data. The fully connected layers at the end of the CNN then interpret these features to classify the text, often outputting a probability score indicating the likelihood of suicidal ideation present in the text. The effectiveness of this approach relies on the quality and diversity of the training data, as well as the complexity and depth of the CNN model used.

*iii) Recurrent Neural Networks (RNN):* Recurrent Neural Networks (RNNs) are a type of neural network particularly suited for processing sequential data, such as time series or text [73]. Unlike traditional neural networks, RNNs have a unique feature: they maintain a form of memory by using their output as input for the subsequent step. This is achieved through loops within the network [74]. In mathematical terms, the hidden state $h_t$ at time $t$ in a simple RNN is calculated as

$$h_t = activation(W_{hh}h_{t-1} + W_{xh}x_t + b_h) \quad (13)$$

where $W_{hh}$ and $W_{xh}$ are weight matrices, $x_t$ is the input at time $t$, $h_{t-1}$ is the previous hidden state, and $b_h$ is the bias. The output $y_t$ is then a function of the current hidden state: $y_t = activation(W_{hy}h_t + b_y$ with $W_{hy}$ being the output layer weights and $b_y$ the output bias.

For suicidal ideation detection in social media text, RNNs are particularly effective due to their ability to process and learn from the sequential nature of language. The process starts with preprocessing the text data, including tokenization and possibly embedding the words into a vector space [8]. The RNN then processes this sequential data, with each word (or token) being input sequentially. The RNN's hidden state updates with each word, effectively allowing the network to remember and utilize the context from previous words. This context understanding is crucial for identifying patterns or linguistic cues indicative of suicidal ideation, such as expressions of despair, hopelessness, or direct mentions of self-harm. The final output layer of the RNN can classify the text based on the learned patterns, potentially indicating whether the text suggests suicidal ideation. The performance of RNNs in this application hinges on the depth of the network, the quality of the training data, and the network's ability to capture and utilize long-term dependencies in the text.

*iv) Long Short-Term Memory (LSTM) and Bidirectional LSTM (BiLSTM):* Long Short-Term Memory (LSTM) networks are an advanced type of Recurrent Neural Network (RNN) designed to better capture long-range dependencies and address the vanishing gradient problem common in standard RNNs [75]. The key to LSTM's effectiveness is its unique cell structure comprising three gates: the input gate $i_t$, the forget gate $f_t$, and the output gate $o_t$. These gates collectively decide what information to retain or discard as the network processes data sequentially. The LSTM cell updates are governed by the following equations:

$$f_t = \sigma(W_f.[h_{t-1}, x_t] + b_f), \text{(Forget Gate)} \quad (14)$$

$$i_t = \sigma(W_i.[h_{t-1}, x_t] + b_i), \text{(Input Gate)} \quad (15)$$

$$\tilde{C}_t = \tanh(W_c.[h_{t-1}, x_t] + b_c), \text{(Cell State Candidate)} \quad (16)$$

$$C_t = f_t * C_{t-1} + i_t * \tilde{C}_t, \text{(Cell State Update)} \quad (17)$$

$$o_t = \sigma(W_o.[h_{t-1}, x_t] + b_o), \text{(Output Gate)} \quad (18)$$

$$h_t = o_t * \tanh(C_t), \text{(Hidden State)} \quad (19)$$

A Bidirectional LSTM (BiLSTM) extends the standard LSTM by introducing a second layer that processes data in the reverse order [76]. While a regular LSTM processes data from past to future (left to right in a sequence), a BiLSTM also processes data from future to past (right to left), effectively capturing information from both directions. This dual processing makes BiLSTMs particularly effective in applications where the context of the entire sequence is crucial for understanding each part of it.

In the context of suicidal ideation detection from social media text, LSTMs excel due to their ability to capture and remember pertinent information across long text sequences [8], [28], which is crucial in understanding the context and nuances of language. By processing sequential data, LSTMs can identify patterns or phrases indicative of suicidal thoughts, such as expressions of hopelessness or mentions of self-harm. In suicidal ideation detection, a BiLSTM would analyze the text from both directions [45], offering a more comprehensive understanding of the context and sentiment expressed, thus potentially improving the accuracy of detecting signs of suicidal ideation.

*v) Transformer models:* Transformer-based models have revolutionized natural language processing by offering a more efficient and effective means of handling sequential data compared to traditional RNNs or LSTMs. The core concept of Transformers is the self-attention mechanism, which computes the output of a layer by attending to all positions within the same layer [77]. The basic equation for self-attention can be written as:

$$Attention(Q, K, V) = softmax\left(\frac{QK^T}{\sqrt{d_k}}\right)V \quad (20)$$

where $Q$, $K$, and $V$ are queries, keys, and values matrices, respectively, and $d_k$ is the dimensionality of the keys. This mechanism allows the model to weigh the importance of different parts of the input sequence differently, making it highly effective for tasks involving contextual understanding.

Different transformer-based models have emerged, each with unique characteristics [78]:

- STATENet integrates spatial-temporal attention networks for enhanced sequence modeling.

- TransformerRNN combines the transformer architecture with RNNs to capitalize on both methods' strengths.

- BERT (Bidirectional Encoder Representations from Transformers) and its variant Sentence-BERT offer deep bidirectional context understanding by pre-training on large text corpora.

- GUSE (Google Universal Sentence Encoder) is optimized for sentence-level embeddings.

- ALBERT (A Lite BERT) is a more efficient version of BERT with shared parameters across layers.

- RoBERTa (Robustly optimized BERT approach) enhances BERT through robust training methods.

- XLNet combines the best of BERT and autoregressive methods for improved language modeling.

In the context of suicidal ideation detection from social media text, these transformer-based models excel due to their ability to understand and interpret the nuances and context of natural language [22]. By analyzing the text data from social media, these models can capture the subtleties of language that indicate suicidal thoughts or tendencies, such as specific phrases, sentiment, and contextual clues. The self-attention mechanism allows the models to focus on relevant parts of the text, potentially identifying warning signs hidden in normal conversation. This can include detecting direct expressions of suicidal thoughts or more subtle indicators like expressions of hopelessness or isolation. The effectiveness of these models in such applications depends significantly on the quality of the training data and the specific architectural choices made in the model design.

*e) Feature engineering*

- Bag of Words (BoW): This model treats text as an unordered collection of words, focusing on the frequency of each word but ignoring the order and context [79]. In suicidal ideation detection, BoW can help identify frequently used words in posts that are indicative of distress or suicidal thoughts, such as "helpless" or "worthless."

- Tokenization: This process involves breaking down text into smaller units, typically words or phrases [80]. For suicidal ideation detection, tokenization is the first step in analyzing social media posts, as it helps in isolating individual words or phrases for further examination.

- Stemming: Stemming reduces words to their root form, often by chopping off prefixes or suffixes. For instance, "running" becomes "run" [81]. In suicidal ideation detection, stemming can help in consolidating different forms of a word to a single root, making it easier to analyze and categorize texts.

- TF-IDF (Term Frequency-Inverse Document Frequency): This is a statistical measure used to evaluate the importance of a word in a document, which is part of a corpus. It increases with the number of times a word appears in the document but is offset by the frequency of the word in the corpus [82]. In suicidal ideation detection, TF-IDF can help in identifying unique terms in individual posts that are unusual in general but common in texts expressing suicidal thoughts.

- N-gram: N-grams are continuous sequences of 'n' items from a given sample of text or speech [83], [84]. For suicidal ideation detection, analyzing bigrams (2-grams) or trigrams (3-grams) can reveal specific phrases that are more indicative of suicidal ideation, such as "end my life".

- Word Embedding: This technique involves mapping words or phrases to vectors of real numbers [85], capturing the context and semantic relationships between words. In suicidal ideation detection, word embeddings can provide a deeper understanding of the context and nuances in social media posts, which simple frequency counts cannot.

- Normalization: This process involves transforming text into a more uniform format, such as converting all characters to lowercase, removing punctuation, or converting numbers to words [86]. In detecting suicidal ideation, normalization ensures that the analysis isn't skewed by superficial variations in the text.

- Lemmatization: Similar to stemming, lemmatization also reduces words to their base or dictionary form, but it does so use linguistic knowledge about the word's proper form or lemma [87]. For instance, "better" is lemmatized to "good". In the context of suicidal ideation detection, lemmatization helps in accurately grouping together different forms of a word, leading to more effective analysis.

Each of these techniques contributes to transforming raw text into a structured, analyzable form, aiding in the identification of language patterns associated with suicidal thoughts and behavior.

*4) Evaluation matrices:* After developing the models, evaluating their performance is crucial. Common evaluation metrics include [88]:

- Accuracy: The ratio of correctly predicted instances to the total instances in the dataset.

- Precision and Recall: Precision measures the proportion of true positive identifications among the positive identifications made by the model, while recall measures the proportion of true positive identifications among the actual positives.

- F1 Score: The harmonic mean of precision and recall, providing a balance between the two metrics.

- AUC-ROC Curve: A performance measurement for classification problems at various thresholds settings.

*5) Challenges to address:* Despite the advancements in methodologies, there are inherent challenges. One major challenge is the balance between model complexity and interpretability. While deep learning models offer advanced capabilities, they often act as 'black boxes', making it difficult to interpret their decision-making processes.

Moreover, the issue of data imbalance, where instances of suicidal ideation are significantly less than non-suicidal posts, can skew model training and affect the reliability of predictions.

Another challenge is the generalizability of these models. Models trained on specific datasets or demographics might not perform effectively when applied to different datasets or populations. This issue underscores the importance of creating diverse and representative training datasets.

The detection of suicidal ideation on social media using machine learning, deep learning, and NLP involves a complex interplay of various methodologies. From dataset collection and preprocessing to the application of advanced algorithms and evaluation, each step plays a crucial role in the effectiveness of the detection process. While significant progress has been made, ongoing research is needed to address the challenges of model interpretability, data imbalance, and generalizability.

## IV. CONCLUSION

Social media has become a valuable data source for detecting mental health conditions, such as suicidal ideation. This review synthesizes current research utilizing machine learning, deep learning, and NLP techniques to identify warning signs in social media posts. The study highlights the advancement of AI in suicide prevention, showcasing models that leverage deep neural networks and transfer learning to decipher complex linguistic patterns indicative of suicidal thoughts. Key findings demonstrate precision rates of 82-97% and recall rates of 71-94% using models such as SVMs, Random Forests, and neural networks. Despite promising results, model robustness must be improved for real-world application.

Challenges include limited, non-representative datasets that hinder generalizability and may introduce demographic biases. Most models were developed using English data from American users, lacking validation across cultures, languages, and demographics. Achieving a balance between accuracy, complexity, and interpretability remains difficult, with simpler models underperforming and complex models being opaque. Future work should focus on testing across age groups and integrating additional risk indicators, alongside addressing privacy concerns and ensuring ethical data use.

Differentiating genuine suicidal intent from rhetorical expressions is an ongoing challenge, as individuals communicate distress uniquely, influenced by personal and cultural factors. Addressing these nuances requires further qualitative research and interdisciplinary collaboration for better contextual understanding. Early identification through improved technology can facilitate timely intervention, but real-world implementation must consider privacy rights and avoid stigmatization. Clear, ethically aligned protocols are essential for the responsible handling of flagged data.

In conclusion, while significant strides have been made in using machine learning for large-scale screening of suicidal ideation, future research should expand datasets, enhance generalizability, and refine contextual interpretation. Ensuring ethical standards and practical protocols will be vital for maximizing the public health benefits of these technological advances.

## ACKNOWLEDGMENT

This research was fully funded by the UMPSA Research Grant Scheme under grant RDU230353 and PGRS2303109.